\documentclass[%
 reprint,
 amsmath,amssymb,
 aps,
]{revtex4-1}

\usepackage{graphicx}
\usepackage{dcolumn}
\usepackage{bm}

\newcommand{\rme}{\mathrm{e}} 
\newcommand{\rmi}{\mathrm{i}}

\usepackage{xcolor}
\usepackage{soul} 
\usepackage{subcaption}
\usepackage{graphics}
\graphicspath{{Figures/}}

\begin{document}

\preprint{APS/123-QED}

\title{Transient perturbative nonlinear responses of plasmonic materials}

\author{Mikko J. Huttunen, Jussi Kelavuori and Marco Ornigotti} 
\address{
    Photonics Laboratory, Physics Unit, Tampere University, FI-33014 Tampere, Finland
}

\date{\today}

\begin{abstract}
Recent investigations on optical nonlinearities of plasmonic materials suggest their responses may be even beyond the usual perturbative description. To better understand these surprisingly strong responses, we develop here a simple but general approach to describe the nonlinear optical response of plasmonic materials up to $n$th perturbation order. We apply the approach to understand spectral broadening occurring in resonant metasurfaces and investigate the enhancement of high-harmonic generation from multiply-resonant metasurfaces, predicting an over million-fold enhancement of higher harmonics.

\begin{description}
\item[DOI]
\end{description}
\end{abstract}

\pacs{Valid PACS appear here}
\maketitle
 
\section{Introduction}
Photonic metamaterials, i.e., artificial materials that enable sub-wavelength length scale control of light, play a key role in modern nanophotonics~\cite{Alu2007, Soukoulis2011}. Besides the study of the linear optical responses of metamaterials, a growing interest has emerged in the recent years to understand also their nonlinear responses~\cite{Kauranen2012review, Lapine2014, ButetReview2015, Li2017, Rahimi2018}, 
as they are essential in many different applications including frequency conversion, high-harmonic generation (HHG),  ultrashort pulse generation and frequency combs~\cite{Brabec2000intense, Kippenberg2011comb}. Spectral broadening and supercontinuum generation (SCG) are particularly interesting nonlinear phenomena that are utilized to make spectrally broad and coherent light sources~\cite{Dudley2006}, which have found numerous applications for example in gas sensing \cite{Langridge2008}, generation of few-cycle pulses \cite{Dudley2004, Foster2005}, optical metrology \cite{Johnson2015}, spectroscopy \cite{Domachuk2008, Petersen2014}, and optical coherence tomography~\cite{Schenkel2003}. Consequently, these two nonlinear phenomena are of considerable scientific and technological interest. 
 
Spectral broadening and SCG occur ubiquitously in solids, liquids and gases~\cite{Alfano1970b, Corkum1986, Chambert2004}. Unfortunately, the intrinsic material nonlinearities giving rise to these phenomena are very weak, making it necessary to use strong excitation pulses from amplified laser systems with peak intensities of the order of $10^{14}$ W/cm$^2$ to exploit those nonlinear effects. Nonlinear optical fibers, for example, provide a relatively flexible platform that can be also utilized with moderate peak intensities~\cite{Dudley2006}. However, their big disadvantage is the requirement for long interaction lengths, which makes them incompatible with the small footprint typical of nanophotonic devices. 

Integrated photonic supercontinuum sources, on the other hand, are characterized by small footprint and compatibility with mass production, representing an ideal platform for the realization of on-chip SCG~\cite{Oh2014, Johnson2015}. The recent demonstrations of SCG from plasmonic nanoparticles and metasurfaces, in particular, have raised the interest to further investigate their properties, in particular the out-of-equilibrium dynamics of conduction electrons occurring in such systems, as they are responsible for both SCG and spectral broadening~\cite{Muhlschlegel2005, Biagioni2012, Chen2018}. Many of the experiments have been performed using only moderate peak intensities ($10^{10}$--10$^{11}$ W/cm$^2$), prompting to investigate the possible origins of the responses, because conventional theory predicts necessity to use orders of magnitude higher intensities to realize efficient SCG from such thin materials~\cite{Chen2018}. 

Moreover, because the complex dynamics of conduction electrons in plasmonic materials dictate their optical properties~\cite{Muhlschlegel2005}, it seems natural to assume that the same dynamics are pivotal also in SCG occurring in such systems. To this aim, the hydrodynamic model has recently re-emerged as a powerful tool to understand the conduction electron dynamics~\cite{Sipe1980, Scalora2010, Krasavin2016}, and has also been used to propose that nonlocal and nonperturbative effects occurring in plasmonic materials could play a major role in the surprisingly strong SCG~\cite{Krasavin2016, Chen2018}. However, it is not yet clear whether other effects could also participate in the process.

In this work, we investigate the transient nonlinear responses of plasmonic systems motivated by the fact that modern experiments are often performed using ultrashort laser pulses. We extend the simple anharmonic oscillator model, commonly used to describe nonlinear responses, by taking into account the transient response of the system and by generalizing the treatment to include up to $n$ perturbation orders. We show that considerable spectral broadening of ultrashort pulses may take place in resonant plasmonic systems, such as in metasurfaces. We also predict that dramatic 6--7 orders of magnitude enhancement of higher-harmonic processes could occur in plasmonic structures exhibiting multiple resonances. 

Our work is organized as follows: in Sect.~II we use the Green function approach to extend the usual anharmonic oscillator model to include the transient dynamics up to $n$th perturbation order. In Sect.~III we discuss how accounting for the transient response of the system could result in a significant broadening of the incident ultrafast pulse. We also discuss how the transient contributions into the light--matter interaction could result in a significant enhancement of the HHG process in multiresonant plasmonic structures. In Sect.~IV, we draw the conclusions. 

\section{Theory} 
We start our analysis by considering the classical anharmonic oscillator model and seek to describe the time-varying displacement of the conduction electron $\tilde{x}(t)$. We assume the electron to be forced into motion by an incident field 
\begin{equation}
\tilde{E}(t) = \frac{1}{\sqrt{2\pi}} \int_{\Delta \omega} E(\omega') \rme^{-\rmi \omega' t} \,d\omega' \,,
\end{equation} 
where the Fourier integral extends over the incident pulse frequencies $\Delta \omega$ centered at the fundamental frequency $\omega$. We further assume $\tilde{E}$ to be strong enough, so that the electron oscillation becomes noticeably anharmonic, while we take $\tilde{E}$ to be weak enough to allow describing the anharmonic motion perturbatively. In this case, the forced oscillation of the electron displacement can be described using the classical anharmonic oscillator model~\cite{BoydBook2003} 
\begin{equation}  \label{Eq:displacement}
	\ddot{\tilde{x}} + 2\gamma \dot{\tilde{x}} +\omega_0^2\tilde{x} +a \tilde{x}^2	= - e \tilde{E}(t) /m \, ,  
\end{equation}
where $\gamma$ is the damping constant, $\omega_0$ is the resonance frequency of the oscillator, $e$ is the electric charge, $m$ is the effective mass of the electron and $a$ is a parameter describing the anharmonicity of the oscillator. 
The tilde and over-dot notations denote time-varying quantities and their time derivatives, respectively. Although no analytical solutions of Eq.~\eqref{Eq:displacement} are available, in the case of weak nonlinearities ($a\tilde{x}\ll \omega_0^2$), an approximate perturbative solution of form
\begin{equation} \label{Eq:xperturb}
    \tilde{x}(t) = \tilde{x}_1(t) +\tilde{x}_2(t) +\tilde{x}_3(t)  + ... + \tilde{x}_n(t)  
    \, ,
\end{equation}
is commonly solved up to the first few orders in terms of its steady-state solution~\cite{BoydBook2003}. Here, we extend the conventional treatment to include the transient response of the oscillator and by finding a general solution taking into account perturbative corrections up to $n$th order. This allows us to calculate the perturbative behavior of the anharmonic oscillator for arbitrary input profiles $\tilde{E}(t)$. 

Substituting the ansatz above into Eq.~\eqref{Eq:displacement} and equating the terms of the same perturbative order, we get the following set of coupled differential equations, for the various perturbation orders:
\begin{subequations}
\label{Eq:perturbdisplacement}
\begin{align}  
	\ddot{\tilde{x}}_1 + 2\gamma \dot{\tilde{x}}_1 +\omega_0^2\tilde{x}_1 
    =& - e \tilde{E}(t) / m \,,    \label{Eq:1st-perD} \\
    \ddot{\tilde{x}}_2 + 2\gamma \dot{\tilde{x}}_2 +\omega_0^2\tilde{x}_2  =& -a \tilde{x}_1^2 \, ,      \label{Eq:2nd-perD} \\	
    \ddot{\tilde{x}}_3 + 2\gamma \dot{\tilde{x}}_3 +\omega_0^2\tilde{x}_3  =& -2a \tilde{x}_1 \tilde{x}_2\,,  \label{Eq:3rd-perD} \quad\\
    \vdots& \nonumber \\
     \ddot{\tilde{x}}_n + 2\gamma \dot{\tilde{x}}_n +\omega_0^2\tilde{x}_n  =& -a\sum_{|\alpha|=2} \begin{pmatrix}
           2 \\
           \alpha
         \end{pmatrix} \tilde{x}^\alpha \,. \label{Eq:nth-perD}
\end{align}
\end{subequations}
The right-hand side of the $n$th equation follows from the multinomial theorem, and takes use of the multi-indices $\alpha=(\alpha_1,\alpha_2,...,\alpha_n)$ and $\tilde{x}^\alpha=\tilde{x}^{\alpha_1}_1\tilde{x}^{\alpha_2}_2...\tilde{x}^{\alpha_n}_n$. Note, that $\tilde{x}_3$ can be seen to arise from cascaded nonlinear processes, where light generated by the 2nd-order correction term ($\tilde{x}_2$) gives rise to higher-order terms via process of sum-frequency generation. Similarly, the higher-order terms can be seen to arise from cascaded processes of lower order.

The complete linear response, i.e., the solution of Eq.~\eqref{Eq:1st-perD}, can be written as a convolution of the Green function $\tilde{G}$ 
of the unperturbed oscillator and the incident field $\tilde{E}$~\cite{Byron1992} 
\begin{align}  \label{Eq:1st-order}
	\tilde{x}_1(t) &=  -\frac{e}{m}   \int^{\infty}_{-\infty}  \tilde{G}(t-t')
	\tilde{E}(t') \,dt' = -\frac{e}{m}\tilde{G} \circledast \tilde{E}
   \, . 
\end{align}
In the last equality we have introduced $\circledast$ as a shorthand to indicate the convolution operation. The time-domain Green function of an under-damped oscillator 
is~\cite{Byron1992}
\begin{equation}
    \tilde{G}(t) = \frac{\rme^{-\gamma t}}{\omega_{tr}} \sin\left( \omega_{tr} t \right)\Theta( t)   \, ,
\end{equation}
where $\omega_{tr}=\sqrt{\omega_0^2 -\gamma^2}$ is the natural frequency of the oscillator and $\Theta(t)$ is the step function imposed due to causality of the system. Note, that Eq.~\eqref{Eq:1st-order} contains an important result: while the motion of an electron described by Eq.~\eqref{Eq:1st-order} will, at equilibrium, oscillate following the driving field $\tilde{E}(t)$, on a timescale of the order of $\tau=\gamma^{-1}$, the electron oscillates at its natural frequency $\omega_{tr}$, rather than following the impinging field. If this transient timescale, which for bulk gold and silver, for example, is around 10 fs and 30 fs, respectively~\cite{Johnson1972}, is comparable with the characteristic timescale of the driving pulse, efficient coupling between the transient and driven electron dynamics will take place.

The solution for the second- and third-order correction terms $\tilde{x}_2$ and $\tilde{x}_3$ can then be found by solving Eqs.~(\ref{Eq:2nd-perD}) and (\ref{Eq:3rd-perD}) with the aid of the first-order solution. To do that, it is convenient to work in the frequency domain, where the convolution can be regarded as a simple multiplication, allowing us to rewrite Eq.~\eqref{Eq:1st-order} as
\begin{align}  \label{Eq:1nd-order_ft}
	\hat{x}_1(\omega) &= -\frac{e}{m}
    \hat{G}(\omega) \hat{E}(\omega) \, ,  
\end{align}
where the hat notation indicates that variable is described in the frequency domain. The frequency-domain Green function of the under-damped oscillator $\hat{G}(\omega)$ is explicitly given as
\begin{equation}
    \hat{G}(\omega) = \frac{A_0}{\omega_{tr}}\left( \frac{1}{\gamma+i(\omega-\omega_{tr})}-\frac{1}{\gamma+i(\omega+\omega_{tr})} \right)\, ,
\end{equation}
where the scaling of the spectral amplitude $A_0$ is dictated by the investigated structure. Using Eq.~\eqref{Eq:1nd-order_ft} and the result above, the solutions of Eqs.~\eqref{Eq:2nd-perD}--\eqref{Eq:nth-perD} can be then written in the following compact form
\begin{subequations}
\label{Eq:3nd-order_terms_FT}
\begin{align}  
    \hat{x}_2(\omega) &= -a\,\hat{G}(\omega) \left[ \hat{x}_1(\omega) \circledast \hat{x}_1(\omega) \right]
    \,, \label{Eq:2nd-order} \\ 
	\hat{x}_3(\omega) &= -2a\,\hat{G}(\omega)  \left[ \hat{x}_1(\omega) \circledast \hat{x}_2(\omega) \right] \,, \label{Eq:3rd-order} \\
	& \vdots \nonumber \\
	\hat{x}_n(\omega) &= -a \,\hat{G}(\omega)  \sum_{|\alpha|=2} \begin{pmatrix}
           2 \\
           \alpha
         \end{pmatrix} \hat{x}^\alpha(\omega) \,. \label{Eq:nth-order}
\end{align}
\end{subequations}
Here, the $n$th-order solution is concisely written using multi-index notation $\hat{x}^\alpha=\hat{x}^{\alpha_1}_1 \circledast \hat{x}^{\alpha_2}_2 \circledast...\circledast \hat{x}^{\alpha_n}_n$. Note, that because convolution in the frequency domain results in a sum between the various frequencies involved in the process, $\hat{x}_2(\omega)$ contains terms that oscillate, in addition to the expected frequency $2\omega$, also at frequencies $2\omega_{tr}$, and $\omega+\omega_{tr}$, corresponding, respectively, to the second-harmonic generation centered at the transient frequency, and sum-frequency generation involving the driving frequency and the transient frequency. Similarly, $\hat{x}_3(\omega)$ will contain terms oscillating, in addition to the expected $3\omega$, also at $3\omega_{tr}$, $2\omega+\omega_{tr}$, and $2\omega_{tr}+\omega$.

Equations~\eqref{Eq:2nd-order}--\eqref{Eq:nth-order} are the main result of our work, and allow us a convenient way to find the general solution for the driven anharmonic oscillator system governed by Eq.~\eqref{Eq:displacement} up to the $n$th perturbative order. This general solution includes the possible transient contributions to the overall response, which may be important either when the driving fields are very short pulses ($\sim$fs) or when the system exhibits dynamics occurring at similar/longer time scales. 

Taking into account the transient dynamics of the conduction electrons in plasmonic systems gives rise to the generation of new frequencies and modulation of existing frequency components associated with the incident pulse. Interestingly, we see that the characteristic transient frequency of the system $\omega_{tr}$ dictates the occurrence of the nonlinear processes allowing means to engineer them. For example, one could control the spectral broadening of an incident pulse by designing a system where the center frequency of the incident pulse and the transient frequency $\omega_{tr}$ coincide, where the strongest spectral broadening is expected to occur near $\omega_{tr}$. This spectral broadening mechanism is interesting because the frequency components retain their phase coherence potentially being useful also for ultrashort pulse generation~\cite{Steinmeyer1999, Brabec2000intense}.

\section{Results and discussion}
To validate our findings, we first calculate the emission spectrum of a metasurface consisting of identical plasmonic gold nanoparticles, by using the perturbative solution to Eq.~(\ref{Eq:xperturb}) as found above. The nanoparticles are taken to exhibit a localized plasmon resonance, peaking near 1.54~eV and associated with the relaxation time $\tau=10$~fs similar to bulk gold~\cite{Johnson1972}. The nonlinear coefficient $a$ was taken to be 20~$ \text{m}^{-1}\text{s}^{-2}$, resulting in calculated SHG efficiencies that agree with earlier experiments on gold nanoparticle arrays~\cite{Czaplicki2018}. 

\begin{figure}[ht]
\includegraphics[width=0.98\linewidth]{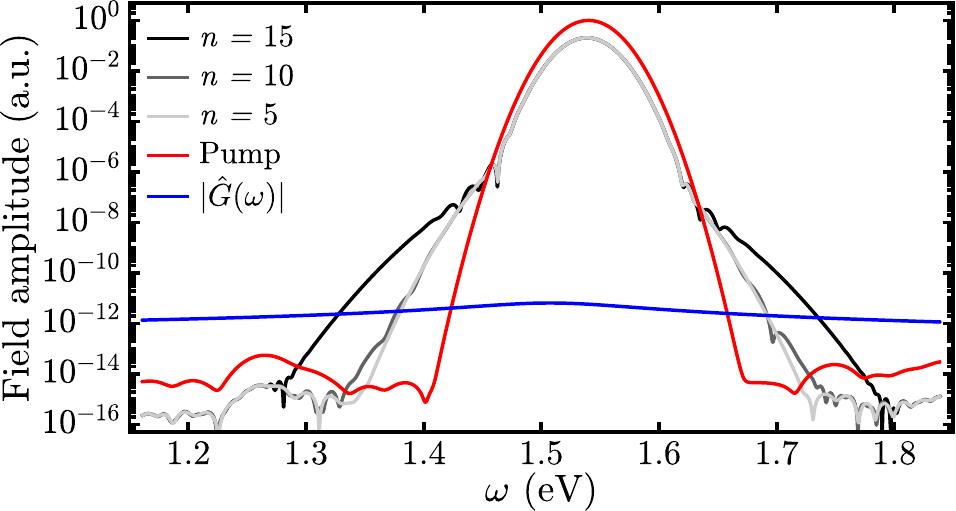} 
\caption{Calculated emission spectrum showing spectral broadening as a function of included perturbation terms $n$. Input pump (red curve) is centered at 805~nm (1.54~eV), and a single material resonance ($\tau=10$~fs) is taken to coincide with the pump wavelength. Blue curve acts as a guide to the eye and visualizes the spectral profile of the Green function associated with the metasurface. 
}
\label{fig:Spectral_broadening}
\end{figure}

We start our analysis by taking an incident pulse with a Gaussian temporal profile with a full width at half maximum of $100$~fs, centered at $\lambda=805$~nm (1.54~eV), and investigate the spectral broadening of the incident field upon interaction. The peak intensity of the incident pulse is assumed to be 20~TW/cm$^2$, which can be readily achieved using amplified systems. A representative calculated emission spectrum obtained from the metasurface defined above is shown in Fig.~\ref{fig:Spectral_broadening}, where the spectral broadening of the incident pump field (red solid curve) is clearly visible. The calculations were repeated while varying the number of included perturbative terms $n=5$, 10, and 15. Surprisingly high perturbative terms ($n>10$) were found to still visibly affect the predicted spectral broadening. This demonstrates the power of the introduced approach, which allows these contributions to be included with ease. Overall, the result shown in Fig.~\ref{fig:Spectral_broadening} suggests that resonant metasurfaces could be used also for spectral broadening of ultrashort pulses. The advantages of metasurfaces compared to traditional materials include compact size, wavelength tunability, and the possibility to engineer the wavefront of interacting free-space beams~\cite{Ye2016, Walter2017}.

\begin{figure*}[t] 
\includegraphics[width=0.95\linewidth]{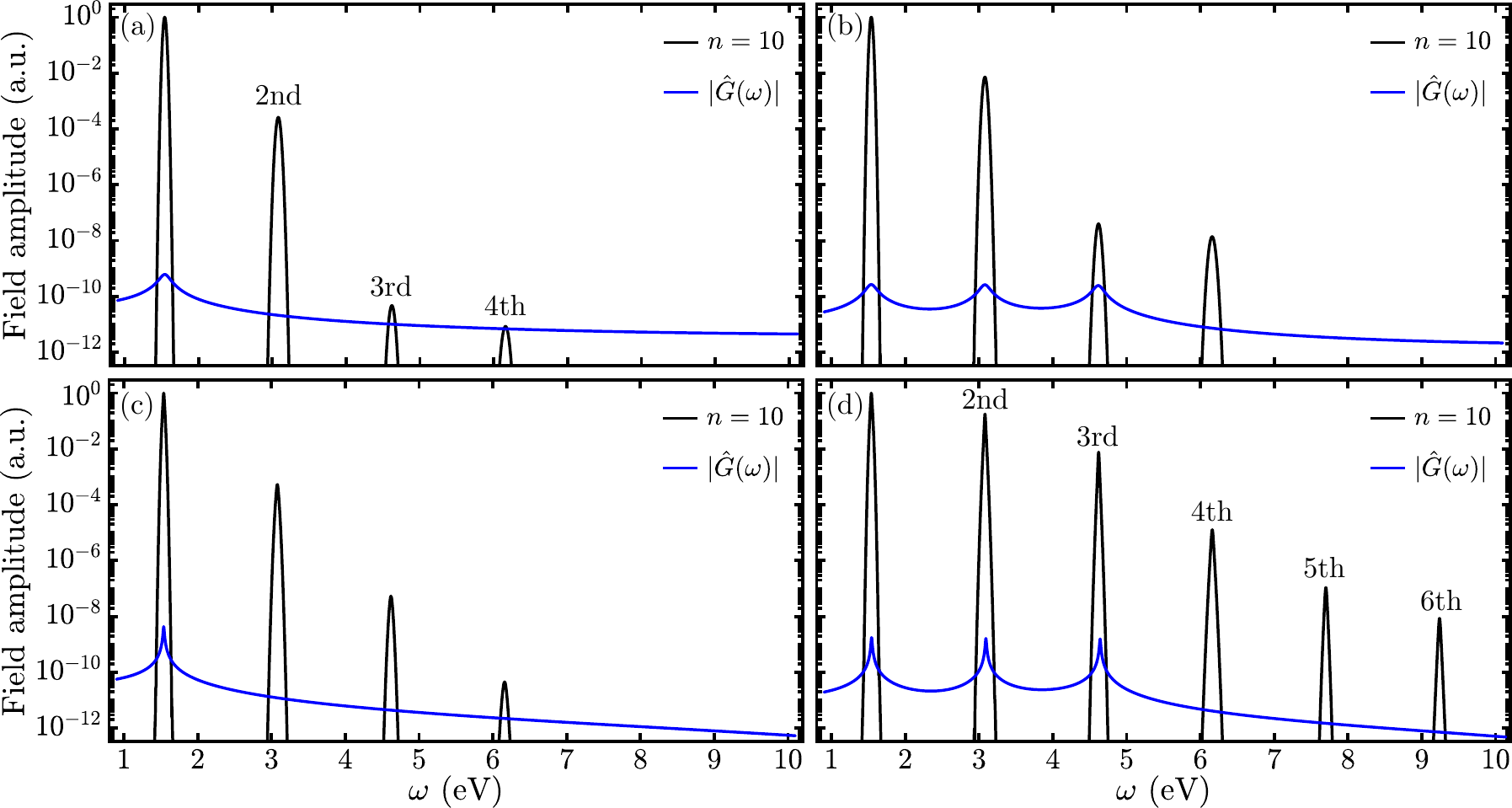}
\caption{Calculated emission spectrum (for $n=10$) using a pump beam centered at 1.54 eV. Blue curves act as guides to the eye and visualize the Green functions associated with the metasurfaces. Emission spectra for singly-resonant (a) and multiresonant (b) metasurfaces associated with resonances exhibiting relaxation times of $\tau=10$~fs. HHG up to the fourth harmonic is visible. Emission spectra for singly-resonant (c) and multiresonant (d) metasurfaces with $\tau=100$~fs. (d) HHG up to the 6th harmonic is clearly visible.}
\label{fig:HHG}
\end{figure*} 

As a second demonstration of the introduced approach, we investigate HHG in plasmonic metasurfaces. It has been recently realized that collective responses of periodic nanostructures can support narrow resonances~\cite{Auguie2008, Huttunen2016a, Kravets2018}. The decay times of these narrow resonances can be considerably longer than those of bulk metals. In addition, the possibility to fabricate metasurfaces supporting multiple resonances has also been recently demonstrated~\cite{Huttunen2019, Reshef2019}. Here, we are interested to understand how the emission spectrum and process of HHG could be engineered by utilizing the above-mentioned plasmonic metasurfaces exhibiting multiple narrow resonances. 

First, we consider the metasurface and incident pulse as described above except for the assumed peak intensity. Here, we use a lower peak intensity of 100~GW/cm$^2$. In the calculations, the first ten perturbative terms are taken into account ($n=10$). The calculated emission spectrum is shown in Fig.~\ref{fig:HHG}(a) demonstrating HHG.
Then we study a multiresonant metasurface that exhibits material resonances at frequencies of 1.54~eV, 3.08~eV, and 4.62~eV, which coincide with the fundamental, second- and third-harmonic peaks of the incident pulse. The relaxation times of these resonances are taken to be the same as above ($\tau=10$~fs). As these resonances occur at the harmonics of the incident pulse frequency (1.54~eV), their presence is expected to enhance HHG. This is also seen in the calculated emission spectrum shown in Fig.~\ref{fig:HHG}(b), where most notably the field amplitude of the fourth-harmonic peak is increased 1560-fold compared to the singly-resonant metasurface [Fig.~\ref{fig:HHG}(a)].

We then investigate singly- and multiply-resonant metasurfaces that exhibit resonances with longer relaxation times ($\tau=100$~fs) than considered above. Such metasurfaces could potentially be realized by utilizing Fabry--P\'erot resonances and collective responses occurring in metasurfaces~\cite{Huttunen2019, Reshef2019}. The calculated emission spectra for these singly- and multiply-resonant metasurfaces are shown in Figs.~\ref{fig:HHG}(c) and ~\ref{fig:HHG}(d), respectively. Comparing the calculated emission spectra of singly-resonant metasurfaces [Figs.~\ref{fig:HHG}(a) and~\ref{fig:HHG}(c)], we see that the narrower resonance results in a 7700-fold (40-fold) enhancement of third-harmonic generation (fourth-harmonic generation). 

It is even more interesting to compare the emission spectra of the singly-resonant metasurface against the multiresonant case with narrow resonances [Figs.~\ref{fig:HHG}(a) and~\ref{fig:HHG}(d)]. In this case, the field enhancements are calculated to be $3.94{\times}10^{6}$ ($10.36{\times}10^{6}$) for the third-harmonic (fourth-harmonic) peak. We also note for the case of the multiresonant metasurface, even the fifth-harmonic and sixth-harmonic peaks are now clearly visible. Together these results prompt to investigate how such multiresonant metasurfaces could be designed and realized~\cite{Huttunen2019,Reshef2019}.

Next, we discuss the origin of the higher-order processes of spectral broadening and HHG, as it might seem surprising that the anharmonic oscillator model [Eq.~\eqref{Eq:displacement}], exhibiting only a quadratic nonlinearity, predicts their occurrence. However, their origin is understood by considering cascaded nonlinear processes, where the lower-order nonlinear processes give rise to effective higher-order nonlinear processes~\cite{Assanto1995, Mu2000, Misoguti2001, Dolgaleva2009, Celebrano2019}. Here, at the beginning of the light--matter interaction, the quadratic nonlinearity gives rise to second-order processes and associated new frequency components [Eq.~\eqref{Eq:2nd-order}], that will later on act as seeds in further quadratic nonlinear interactions [Eq.~\eqref{Eq:3rd-order}]. These and subsequent cascaded second-order processes will then effectively manifest as higher-order processes [Eq.~\eqref{Eq:nth-order}]. Therefore, we see that by using the introduced approach such cascaded higher-order processes can be taken into account simply by including higher-order perturbative terms. This is interesting because our approach allows simple and intuitive means to understand and even engineer the occurrence of higher-order processes in metasurfaces.

Last, we consider the asymptotic behavior of the model by taking an increasingly long relaxation time $\tau$ (i.e. vanishingly small damping term $\gamma$). In this case, the transient responses no longer quickly decay suggesting that they remain non-negligible even under continuous-wave illumination. Despite vanishingly small $\gamma$ is strictly non-physical, we note that recent experimental and numerical investigations on collective responses of metal nanoparticle arrays, known as surface lattice resonances, suggest that systems associated with quite long relaxation times (${\sim}$ps) can be fabricated~\cite{Auguie2008, Huttunen2016a, Bin-Alam2020}.        

\section{Conclusions}
We have investigated the nonlinear optical responses of plasmonic materials by extending the anharmonic oscillator model, commonly used to describe nonlinear responses, by taking into account also the transient response of the system and by finding a general solution including up to $n$ perturbative orders. The approach allows us to better understand the nonlinear responses of these materials and provides means to engineer their responses for applications, such as for spectral broadening or frequency conversion. Our results suggest, that considerable spectral broadening of ultrashort pulses may occur in metasurfaces due to the transient response of the system. We also predict that higher-harmonic processes could be enhanced over a million-fold by utilizing multiply-resonant plasmonic structures. 

\section*{Acknowledgements}
We thank Ksenia Dolgaleva for her valuable feedback and comments. We acknowledge the support of the Academy of Finland (Grant No. 308596) and the Flagship of Photonics Research and Innovation (PREIN) funded by the Academy of Finland (Grant No. 320165). 

\bibliography{NLOmetasurfaces}

\end{document}